\documentclass[reprint,aps,prl]{revtex4-1}
\usepackage{amsmath,graphicx,multirow}

\begin{document}

\title{Measurement of nonlinear piezoelectric coefficients using a micromechanical resonator}
\date{\today}

\author{Joseph A. Boales}
\author{Shyamsunder Erramilli}
\author{Pritiraj Mohanty}
\affiliation{Department of Physics, Boston University, 590 Commonwealth Avenue, Boston, MA 02215, US}

\begin{abstract}
We describe and demonstrate a method by which the nonlinear piezoelectric properties of a piezoelectric material may be measured by detecting the force that it applies on a suspended micromechanical resonator at one of its mechanical resonance frequencies. Resonators are used in countless applications; this method could provide a means for better-characterizing material behaviors within real MEMS devices. Further, special devices can be designed to probe this nonlinear behavior at specific frequencies with enhanced signal sizes. The resonators used for this experiment are actuated using a 1-$\mu$m-thick layer of aluminum nitride. When driven at large amplitudes, the piezoelectric layer generates harmonics, which are measurable in the response of the resonator. In this experiment, we measured the second-order piezoelectric coefficient of aluminum nitride to be $-(23.1\pm14.1)\times10^{-22}\ \mathrm{m/V^2}$.
\end{abstract}

\maketitle

Nonlinear materials and components are used in virtually all modern devices, and nonlinear effects can be found in every branch of science. Engineers often take great pains to avoid such nonlinearities, but they are nonetheless unavoidable---and often helpful---in many devices. In electronics, transistors and diodes are some of the most important components, yet they exhibit interesting nonlinear behavior that is useful for applications such as signal rectification and amplification \cite{Horowitz1989,Read1963}. Optical scientists and engineers often take advantage of nonlinearity for applications like parametric amplification, optical rectification, and harmonic frequency generation, among others \cite{Boyd2008}. Dating back to the time of Newton, nonlinearity has played a role in fundamental physics with the discovery of Newton's Law of Universal Gravitation \cite{newton1729experimental}. The nonlinearity in this law is even relevent today for objects such as satellites; the drag and orbital period are nonlinear in orbital radius, so small variations in the orbital radius can lead to significant variations in orbital motion \cite{Subbarao2008}.

One application in which nonlinearity is becoming increasingly important is in piezoelectric thin-film devices such as resonators \cite{Li2017} and transducers \cite{Liu2015,Benjeddou2018,Sinha2009}. In the piezoelectric effect, an applied mechanical stress produces an electric field; in the inverse effect, an applied electric field produces a strain. Piezoelectric materials provide a convenient way to convert between mechanical displacements or vibrations and electrical signals. As such, there are frequent publications on using piezoelectrics for energy harvesting \cite{Howells2009,Toprak2014}, communication \cite{Boales2017,Satoh2005}, force measurements \cite{Boales2017a,Tansock1992,Butt2005}, and high-precision motion \cite{Zhong2008,Xu2017}, among other applications. In many of these applications, it is desirable to reduce the size of every component of the system as much as possible. However, as components become smaller, nonlinearity becomes ever more important. For highly sensitive applications, it is possible that the unaccounted-for presence of nonlinearity can affect device performance, or that the nonlinearity itself may be useful in further enhancing a device's capabilities.

In the case of piezoelectric materials, if a large electric field is applied, its strain may no longer vary linearly, which can lead to nonlinear phenomena such as second harmonic generation (SHG) \cite{Miller1964,Umeda1998}. Despite the importance of this type of nonlinearity, the methods for measuring the nonlinear components of the piezoelectric coefficients remain limited by various aspects of the measurement, and, based on our literature searches, much of the focus has been in using lead zirconate titanate (PZT) \cite{ISHII2003,Royston1998}. Many current methods use a thin-film device that was fabricated specifically for the test \cite{ISHII2003,Feld2010}; however, such methods may not be able to account for process-dependent changes in the piezoelectric material that may occur during manufacturing of a real device, such as residual stresses that may be present after the removal of a sacrificial oxide.

Here, we show that it is possible to quantify the nonlinear piezoelectric properties by measuring the spectral behavior of a suspended micromechanical resonator at various electric field strengths. This method has the advantage that it can be used at any frequency at which the resonator can vibrate. Further, it can be performed using resonators that are fabricated for real systems or applications.

Microelectromechanical systems (MEMS) resonators can be designed to have a wide operating range in the frequency spectrum or to have high-quality factor resonances. When driven at these resonances, the output signal of the devices can be as much as 10 million times larger than signals produced by off-resonance driving \cite{Ghaffari2013}. By taking advantage of this signal enhancement behavior, resonators can be designed and used to enhance many different types of driving forces.

Piezoelectric materials provide a convenient method for applying highly-repeatable forces to the resonators. If a time-varying signal voltage is applied to the piezoelectric, the force that is applied to the resonator is generated by the center-of-mass (COM) motion of the piezoelectric layer. The magnitude of the applied force can be estimated beginning with the piezoelectric constitutive relations. Using the Einstein summation convention, these are \cite{Ikeda1990}, 
\begin{subequations}
\begin{align}
S_{ij} &= s_{ijkl}^E T_{kl} + d_{kij} E_k \\
D_i &= d_{ikl} T_{kl} + \varepsilon_{ik}^T E_k,
\end{align}
\end{subequations}
where $S$ is the strain, $s^E$ relates to the elastic modulus at constant electric field, $T$ is the stress, $d$ is the piezoelectric coefficient, $E$ is the electric field, $D$ is the electric displacement field, $\varepsilon^T$ is the electric permittivity at constant stress, and subscripts $i$, $j$, $k$, and $l$ denote Cartesian coordinate directions where 1 is along $x$, 2 is along $y$, and 3 is along $z$. To a good approximation, there is negligible stress in the driving electrode and strain in the plane of the piezoelectric film is zero. The force that the piezoelectric applies to the silicon structure can be calculated as,
\begin{equation}
F(t) = -(m_e + m_\mathrm{AlN}) \ddot{y},
\label{eq:Newton}
\end{equation}
from which we obtain
\begin{equation}
F(t) = -\left( \frac{1}{2} m_\mathrm{AlN} + m_e\right) t_0 \ddot{S}_z.
\end{equation}
where $m_\mathrm{AlN}$ is the mass of the AlN between the electrodes, $m_e$ is the mass of the top electrode, $t_0$ is the un-strained thickness of the AlN, and $S_z$ is the strain perpendicular to the film plane. In order to verify the accuracy of this forcing equation to linear order, we used COMSOL Multiphysics, a finite element simulation program, to analyze a simple 1-micron cube of aluminum nitride under single-frequency excitation potential and measured the reaction force at the base of the material.

For a nonlinear piezoelectric coefficient, we can write (neglecting cross-terms)
\begin{equation}
S_z \approx \sum\limits_n d_{33}^{(n)} E_z^n = d_{33}^{(1)} E_z + d_{33}^{(2)} E_z^2 + ...
\label{eq:strain}
\end{equation}
where we adopt the notation that exponents in parentheses denote the order of linearity, and exponents not in parentheses denote powers. For example, $d_{33}^{(n)}$ is the $n$th-order piezoelectric coefficient for strains in the $z$-direction given an electric field along the $z$-axis.

For single-frequency sinusoidal driving, $E_z = (V_0/t_0) \sin \left( \omega t \right)$ where $V_0$ is the amplitude of electric potential difference applied across the piezoelectric material, $\omega$ is the driving angular frequency, and $t$ is time. For this applied electric field, the force applied to the resonator can be calculated by using the strain in equation (\ref{eq:strain}) in equation (\ref{eq:Newton}), then taking the appropriate derivatives to obtain
\begin{align}
F(t) = \left( \frac{1}{2} m_\mathrm{AlN} + m_e \right) &t_0 \omega^2 \nonumber \\
\times \sum\limits_{n=1} \frac{d_{33}^{(n)} V_0^n}{t_0^n} &[ -n \sin^n \omega t \nonumber \\
+ &n (n-1) \sin^{n-2} \omega t \cos^2 \omega t ].
\label{eq:nonlinear_force}
\end{align}

At each order of nonlinearity $n$, this expression can be expanded in linear powers of sines and cosines that are spectrally located at integer multiples of the driving frequency. Further, for each value of $n$, the nonlinear forcing signal is only present at the $n$th harmonic and lower harmonics of the same parity (i.e., only at even or only at odd lower harmonics); the size of the nonlinearity's contribution is proportional to $V_0^n$.

For complex resonators such as the ones used in our experiments, it is often convenient to one-dimensionalize their motion. This can be done using the equation of motion for a damped harmonic oscillator,
\begin{equation}
\ddot{z} + \frac{\omega_0}{Q} \dot{z} + \omega_0^2 z = \frac{F(t)}{m}
\label{eq:eom}
\end{equation}
where $z$ is the effective position of the resonator, $Q$ is the quality factor of the resonance mode, $\omega_0$ is the resonance frequency, $F(t)$ is the driving force, and $m$ is the effective mass of the resonator mode. Taking the Fourier transform of equation (\ref{eq:eom}) and solving for frequency-dependent magnitude of the resonator's response, $|\tilde{Z}(\omega)|$, gives
\begin{equation}
\left| \tilde{Z}(\omega) \right| = \frac{\left|\tilde{F}(\omega)\right|}{m} \frac{1}{\sqrt{\left(\omega_0^2 - \omega^2\right)^2 + \left(\frac{\omega_0 \omega}{Q}\right)^2}}.
\label{eq:response}
\end{equation}
where $|\tilde{F}(\omega)|$ is the magnitude of the Fourier-transformed applied force from equation (\ref{eq:nonlinear_force}).

\begin{figure}
\centering
\includegraphics[width=\columnwidth]{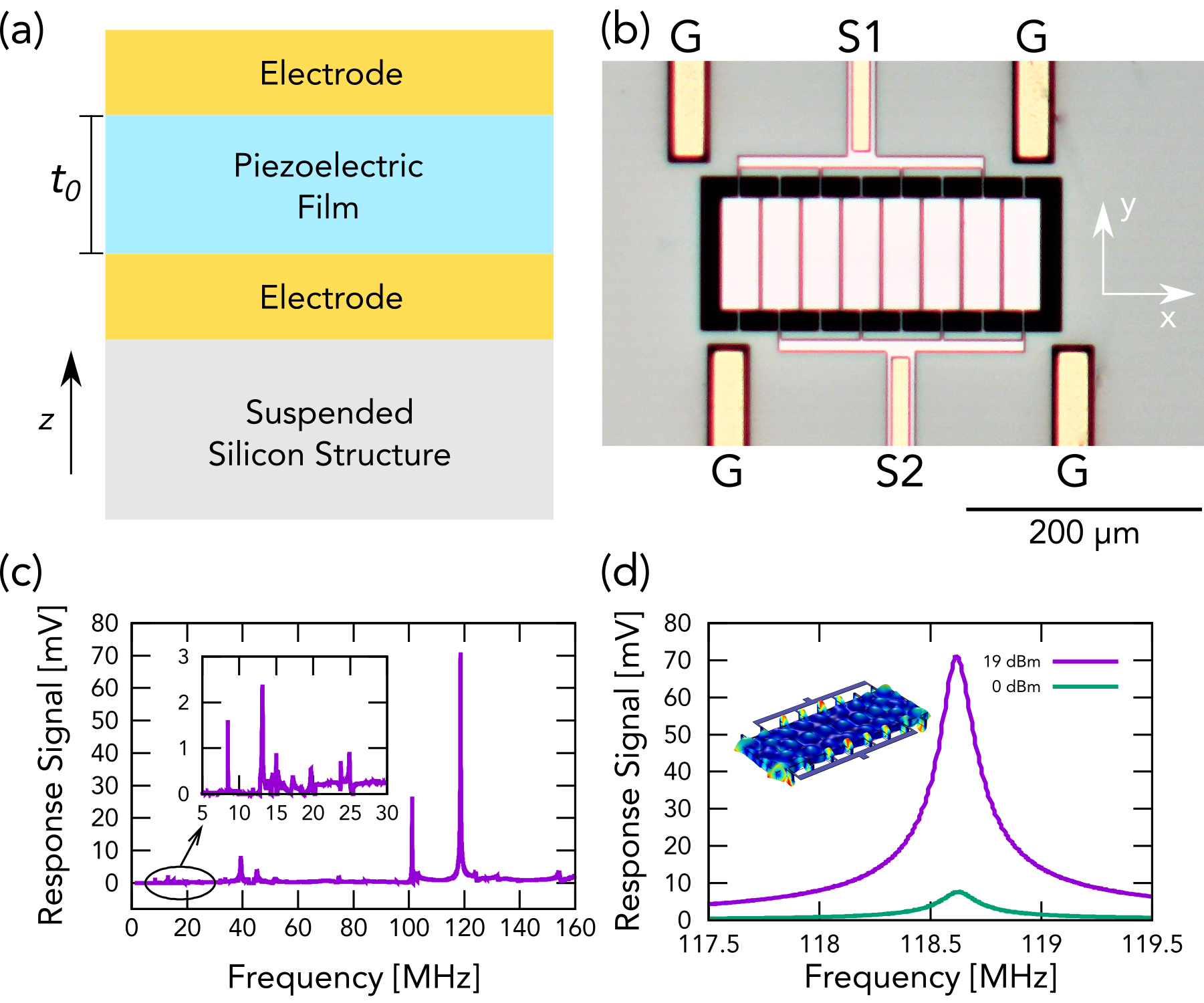}
\caption{\textbf{(a)} Layer structure of resonators. \textbf{(b)} Micrograph of resonator used in experiments. \textbf{(c)} Frequency spectrum of resonator from 1 to 160 MHz for 19-dBm driving power. The frequency spectrum between 5 and 30 MHz is magnified to illustrate the rich spectral behavior of these resonators. \textbf{(d)} Shape of primary resonance peak for 0- and 19-dBm driving with inset modeshape.}
\label{fig:fig1}
\end{figure}

As illustrated in Figure \ref{fig:fig1}(a), the resonators in our experiments are constructed (from bottom to top layer) of a 10-$\mathrm{\mu}$m silicon structure, a 200-nm molybdenum grounding electrode, a 1-$\mathrm{\mu}$m aluminum nitride active layer, and 200-nm signal electrodes. While there are virtually limitless device geometries, for this experiment, we use a rectangular plate-type resonator with side lengths 272 $\mu$m and 96 $\mu$m, and sixteen 3-$\mu$m-by-15-$\mu$m anchors. A micrograph of one of the resonators used for this experiment is shown in Figure \ref{fig:fig1}(b). In this resonator, the top layer consists of two sets of interdigitated electrodes (IDTs).

A signal can be applied to one set of the IDTs via the pad labeled ``S1" or ``S2", and the response of the resonator can be detected using the other. The applied signal applies a force to the suspended silicon structure via the inverse piezoelectric effect, causing it to vibrate as described above; the response of the resonator (i.e., the amplitude of its vibration) is measured at the other electrodes via the direct piezoelectric effect. The pads labeled ``G" are used as ground references. 

In equation (\ref{eq:response}), the amplitude $|\tilde{Z}(\omega)|$ describes the contribution to the output signal that results from deformation of the resonator structure. In general, there will also be a signal contribution that results from the center-of-mass acceleration of the output transducer, which is proportional to the the second time derivative of the deflection, or $\omega^2 |\tilde{Z}(\omega)|$.

Before attempting to quantify the resonator's nonlinear piezoelectric coefficients, we characterized the spectral behavior of the resonator. The resonator's response between 1 and 160 MHz is shown in Figure \ref{fig:fig1}(c). The resonator's largest resonance occurs for a driving frequency of 118.626 MHz. This resonance is shown in Figure \ref{fig:fig1}(d), and its corresponding mode shape (calculated using COMSOL Multiphysics) is shown in the subset.

When a sufficiently large potential is applied to the piezoelectric layer at frequency $f$ (where $f=\omega / 2 \pi$), the layer will generate significant additional forces at integer multiples of that frequencies (which will be measurable as long as they exceed the effective thermal Johnson forcing), as described by equation (\ref{eq:nonlinear_force}). From our analysis, it is evident that all of the nonlinear contributions can be determined by doing separate first and second harmonic signal measurements for the same measurement frequency. That is, by driving and measuring at frequency $f$ (we use the shorthand ``D:f,M:f"), then separately driving at frequency $f/2$ and measuring at frequency $f$ (``D:f/2,M:f").

\begin{figure}
\centering
\includegraphics[width=\columnwidth]{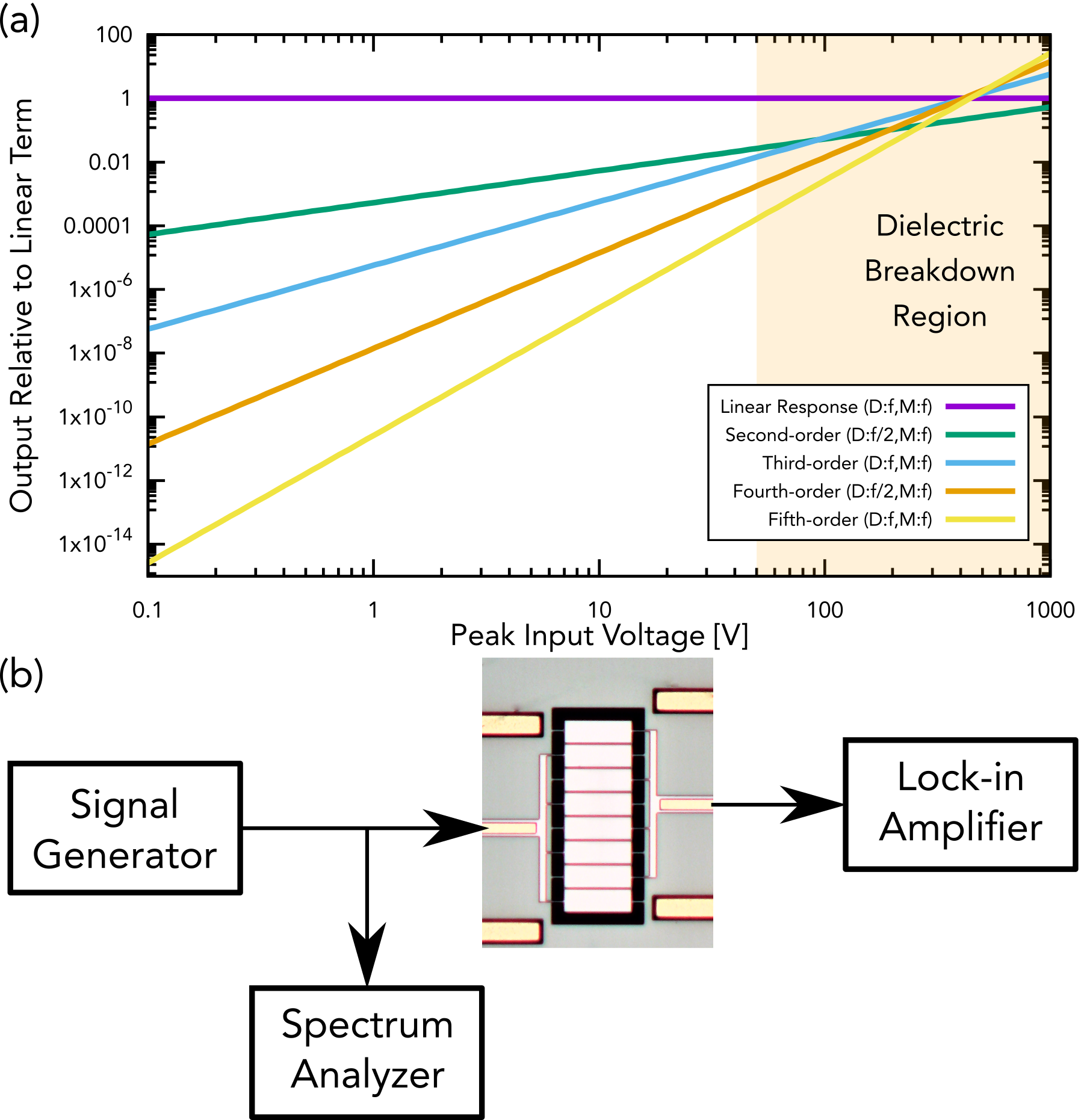}
\caption{\textbf{(a)} Based on published values for quartz, these are the expected relative signal sizes compared to the linear response. In the key, ``D" denotes the driving frequency, and ``M" denotes the measurement frequency. \textbf{(b)} Schematic diagram of the experiment. A signal generator supplies a potential to a micromechanical resonator. The input and output signals are simultaneously measured using a spectrum analyzer and lock-in amplifier, respectively.}
\label{fig:fig2}
\end{figure}

An experiment would require measurements over a range of driving amplitudes. Figure \ref{fig:fig2}(a) shows, using the nonlinear values for quartz \cite{Besson1974}, how the relative signal contributions may vary as a function of driving amplitude. For our devices, the dielectric breakdown of AlN occurs at a maximum of 50 to 100-volts, depending on the excitation frequency  \cite{Ruemenapp1999}. At that amplitude, the second- and third-order contributions could be as large as 1\% of the linear response magnitude.

In order to make these measurements, we designed the experiment described by the schematic in Figure \ref{fig:fig2}(b). In this experiment, the resonator is driven using a Rohde \& Schwarz SMY01 signal generator, and the input and output voltages are simultaneously measured using an Agilent N9000A spectrum analyzer and a Stanford Research Systems SR844 lock-in amplifier, respectively. The input signal measurement is necessary in obtaining the true value of $V_0$.

The output signal is measured at frequency $f$ while driving at frequency $f$, then separately while driving at frequency $f/2$. Simultaneously, the input signal is measured at both frequencies $f$ and $f/2$. This allows us the quantify the sizes of the harmonic distortions from the signal generator and calibrate them out.

After measuring the data, we obtain calibration fits to the ``D:f,M:f" data via a polynomial in odd powers of $V_0$, 
\begin{equation}
g_1(V_0) = a_1 V_0 + a_3 V_0^3.
\label{eq:odd_powers}
\end{equation}
Next, using this calibration with the measured second harmonic produced by the signal generator, we remove the second harmonic signal contribution from the output ``D:f/2,M:f" data. Measurements of the remaining signal are used to obtain the second-order nonlinear piezoelectric coefficient described in this text by fitting it to a polynomial in even powers of $V_0$,
\begin{equation}
g_2(V_0) = a_0 + a_2 V_0^2 + a_4 V_0^4.
\label{eq:even_powers}
\end{equation}

\begin{figure}
\centering
\includegraphics[width=\columnwidth]{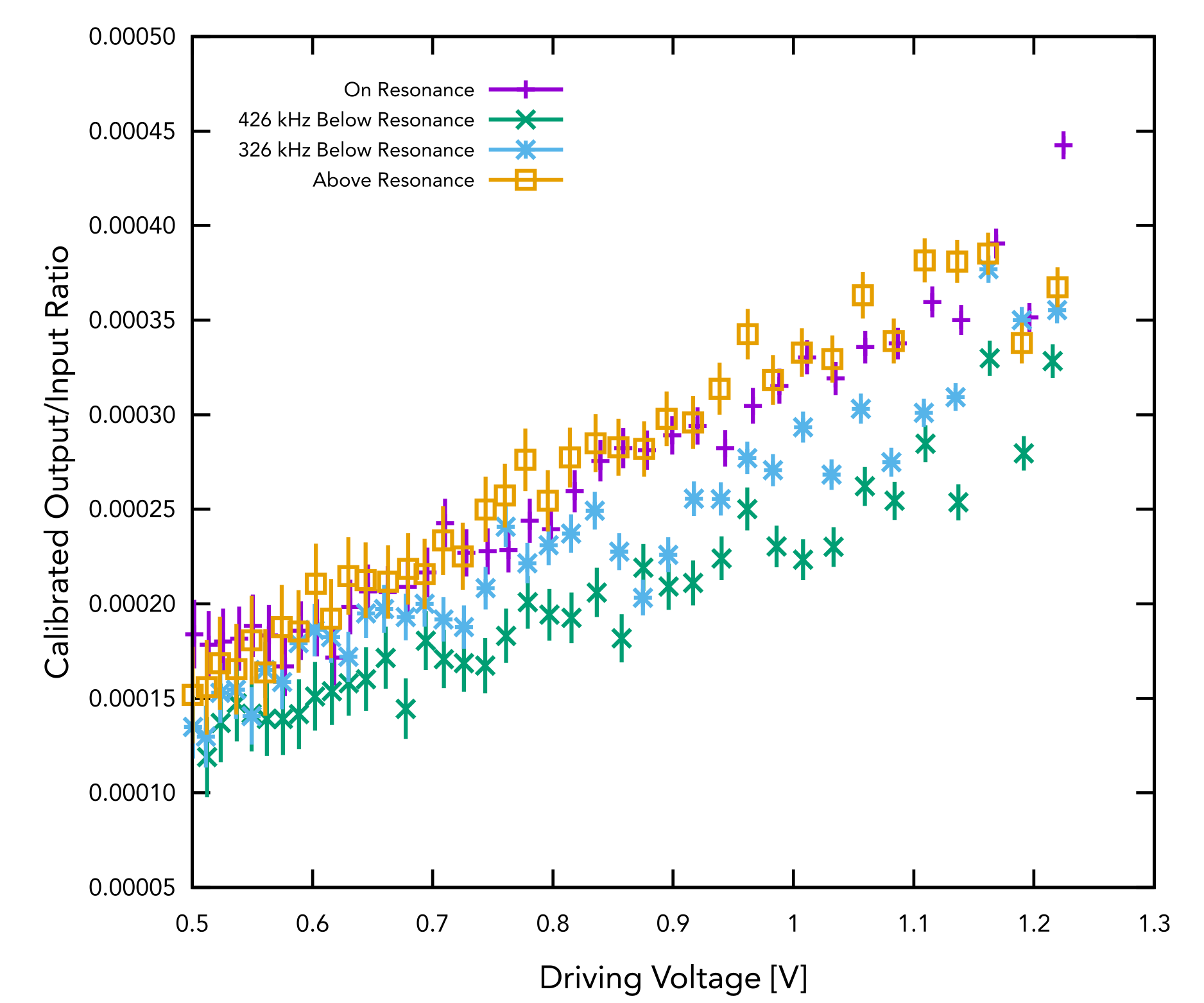}
\caption{Plot of the ratio of the magnitude of the calibrated output signal measured at the second harmonic to the input driving signal as a function of driving signal. Using this and more data, we obtained estimates of the second-order nonlinear piezoelectric coefficient.}
\label{fig:fig3}
\end{figure}

Figures \ref{fig:fig3}(a-b) show sample data that was collected near one of the mechanical resonances using this method. We performed measurements at various frequencies using multiple devices of the same design. Most measurements were taken within 5\% of the primary resonance. As may be expected, the resonance frequency varied between resonators by as much as 4 MHz. In order to obtain quantitative values for the nonlinear piezoelectric coefficients, the coefficients $a_n$ (for integer $n$) in equations (\ref{eq:odd_powers}) and (\ref{eq:even_powers}) can be compared to $a_1$. The second-order term can be determined using the function
\begin{equation}
d_{33}^{(2)} = 2 d_{33}^{(1)} t_0 \frac{a_2}{a_1}
\end{equation}
Equations (\ref{eq:odd_powers}) and (\ref{eq:even_powers}) can be further expanded in higher odd and even terms, respectively, and similar expressions of the form
\begin{equation}
d_{33}^{(n)} = c_n d_{33}^{(1)} t_0^{n-1} \frac{a_n}{a_1},
\end{equation}
where $c_n$ is a harmonic-dependent coefficient determined using equation (\ref{eq:nonlinear_force}), can be used to obtain higher-order nonlinear piezoelectric coefficients.

During this experiment, we measured the second-order piezoelectric coefficient of aluminum nitride using two separate devices near their primary resonances. For the first device, we measured the constant to be $-(26.3\pm10.5)\times10^{-22}\ \mathrm{m/V^2}$; and for the second device, we measured the constant to be $-(17.7\pm15.2)\times10^{-22}\ \mathrm{m/V^2}$. Combining all measurements for both devices, we estimate the constant to be $-(23.1\pm14.1)\times10^{-22}\ \mathrm{m/V^2}$.

Though second-order \textit{dielectric} nonlinearities in AlN have been well-studied \cite{Guo2016,Xiong2012}, these are the first published measurements of the second-order nonlinear piezoelectric coefficients of AlN. We have used quartz as a material for comparison, and found that our measured value is comparable to its value. Quartz has a linear piezoelectric coefficient of about 2.2 pm/V, and a second-order nonlinear constant of $(-17\ \mathrm{to}\ -40)\times10^{-22}\ \mathrm{m/V^2}$ \cite{Besson1974}.

While this method works best for measurements of the nonlinear piezoelectric coefficients at the suspended structure's mechanical resonance, off-resonance measurements are also possible. However, we find that, with our equipment used off-resonance, there is a significantly higher measurement uncertainty. In order to improve this method, we suggest using additional filtering on the input line to reduce the sizes of the harmonic distortions produced by the signal generator, such as the one used by Tashiro et. al \cite{TASHIRO2002}. This can reduce or eliminate the need for removing the second-order and higher harmonic contributions from the driving signal. In addition, low-noise pre-amplifiers and higher-precision signal measurement equipment should improve the results.

We have successfully shown that it is possible to measure second- and higher-order nonlinear coefficients of piezoelectric materials my measuring the force the material applies to a micromechanical resonator. This method may be useful in characterizing the effects of fabrication processes and phenomena on the piezoelectric materials used in devices for real-world applications.
\bibliography{references}
\end{document}